\documentclass[%
reprint,
amsmath,
amssymb,
aps,
pra,
]{revtex4-2}

\usepackage{graphicx}% Include figure files
\usepackage{dcolumn}% Align table columns on decimal point
\usepackage{bm}% bold math
\begin{document}

\preprint{APS/123-QED}

\title{Neutron ghost imaging}% Force line breaks with \\

\author{Andrew M. Kingston}
 \affiliation{Department of Applied Mathematics, Research School of Physics and Engineering, The Australian National University, Canberra, ACT 2601, Australia}
 \affiliation{CTLab: National Laboratory for Micro Computed-Tomography, Advanced Imaging Precinct, The Australian National University, Canberra, ACT 2601, Australia}

\author{Glenn R. Myers}
 \affiliation{Department of Applied Mathematics, Research School of Physics and Engineering, The Australian National University, Canberra, ACT 2601, Australia}
 \affiliation{CTLab: National Laboratory for Micro Computed-Tomography, Advanced Imaging Precinct, The Australian National University, Canberra, ACT 2601, Australia}

\author{Daniele Pelliccia}
\affiliation{Instruments \& Data Tools Pty Ltd, Victoria 3178, Australia}

\author{Filomena Salvemini}
\affiliation{Australian Centre for Neutron Scattering, Australian Nuclear Science and Technology Organisation, New Illawarra Rd, Lucas Heights, NSW 2234, Australia}

\author{Joseph J. Bevitt}
\affiliation{Australian Centre for Neutron Scattering, Australian Nuclear Science and Technology Organisation, New Illawarra Rd, Lucas Heights, NSW 2234, Australia}

\author{Ulf Garbe}
\affiliation{Australian Centre for Neutron Scattering, Australian Nuclear Science and Technology Organisation, New Illawarra Rd, Lucas Heights, NSW 2234, Australia}

\author{David M. Paganin}
\affiliation{School of Physics and Astronomy, Monash University, VIC 3800, Australia}

\date{\today}% It is always \today, today,

\begin{abstract}
Ghost imaging is demonstrated using a poly-energetic reactor source of thermal neutrons. The method presented enables position resolution to be incorporated, into a variety of neutron instruments that are not position resolving. In an imaging context, ghost imaging can be beneficial for dose reduction and resolution enhancement. We also demonstrate a super-resolution variant of the method, namely a parallel form of neutron ghost imaging, with the ability to significantly increase the spatial resolution of a pixelated detector such as a CCD or CMOS camera. Extensions of our neutron ghost-imaging protocol are discussed in detail and include neutron ghost tomography, neutron ghost microscopy, dark-field neutron ghost imaging, and isotope-resolved {\it color} neutron ghost imaging via prompt% and delayed%
-gamma-ray bucket detection.
\end{abstract}

\maketitle

\section{Introduction}

Ghost imaging was originally developed in the setting of visible-light quantum optics \cite{Klyshko1988,Belinskii1994,Pittman1995optical,Strekalov1995}. The {\it spooky} action at a distance of quantum entangled photons (initially thought to be required for the technique) gave rise to its name. It was later determined that only the correlation property of the photons is required \cite{bennink2002two, erkmen2008unified} and classical forms of ghost imaging have since been developed \cite{Erkmen2010}. Ghost imaging has the ability to enhance signal-to-noise ratio (SNR) \cite{MultiplexAdvantageGI}, and reduce dose given significant {\it a posteriori} knowledge of the object \cite{katz2009compressive}. The {\it classical} ghost imaging variant works as follows: An ensemble of spatially random illuminating patterns strikes a beamsplitter, with this first beam having its intensity distribution recorded using a position sensitive detector; the second beam passes through a sample of interest and then has its total transmitted intensity recorded using a large single-pixel detector called a ``bucket,'' $B$.  The position-sensitive detector, $P$, records images that contain no information about the object, since none of the imaging quanta (photons, neutrons, electrons, etc.)---registered by the pixels of $P$---have ever interacted with the object.  Conversely, imaging quanta that are registered by the bucket detector have passed through the object, but such quanta are never measured with a position sensitive detector.  While neither of the signals at $B$ or $P$ individually contain position-sensitive information regarding the sample, a ghost image of the sample may be reconstructed via intensity--intensity correlations between the signals at $B$ and $P$ \cite{katz2009compressive,Bromberg2009ghost}.  This may be viewed as a parallel version of the intensity--intensity correlation experiment of Hanbury Brown and Twiss \cite{HBT1,HBT2,HBT3}.

\begin{figure}
    \centering
    \includegraphics[width=1.0\linewidth,trim={0 1cm 0 0},clip]{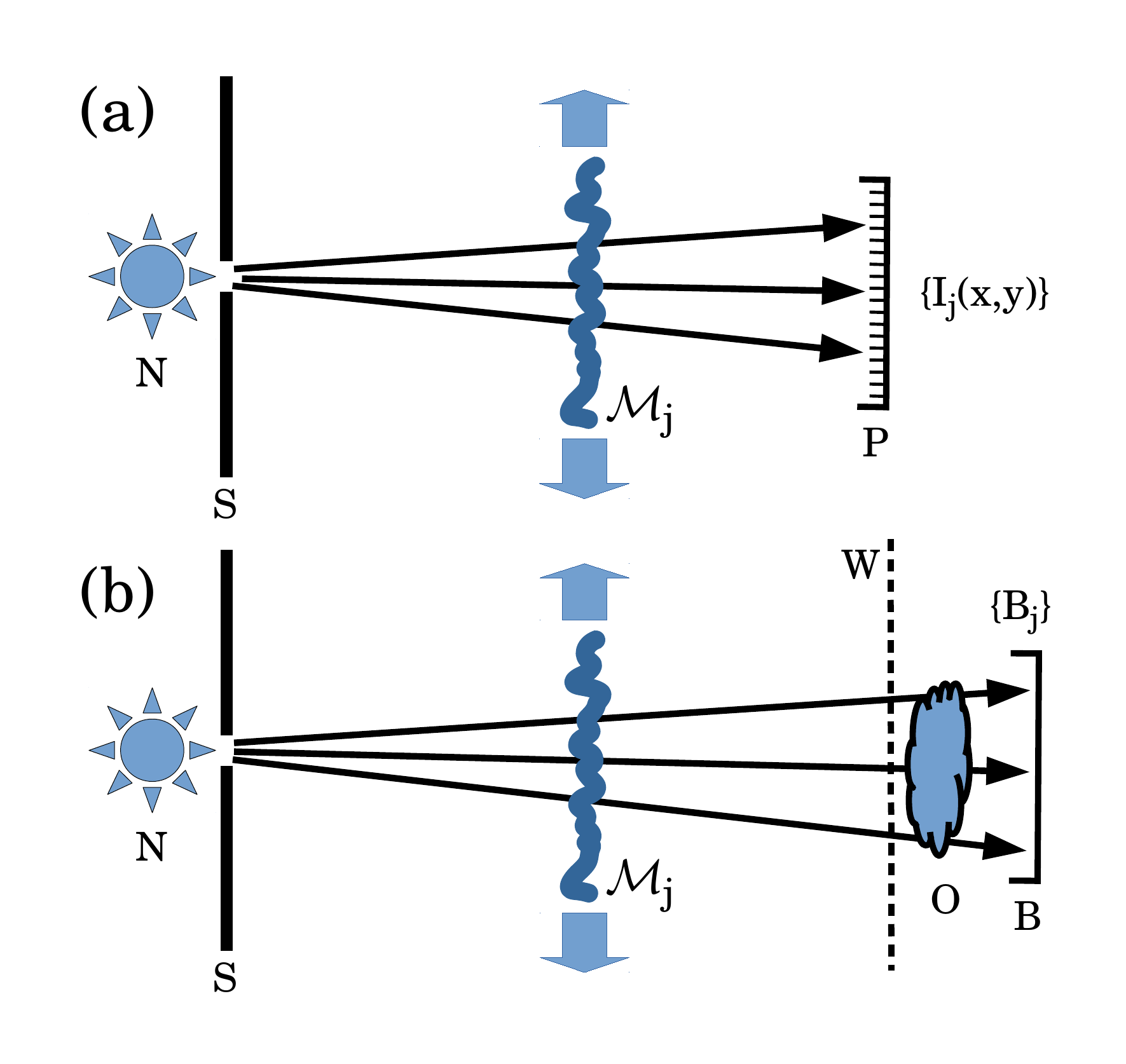}
    \caption{Setup for computational transmission ghost imaging. (a) Imaging quanta (e.g.~visible-light photons, neutrons, x-rays, gamma rays etc.) from a source $\Sigma$ pass through a slit $S$ before traversing a spatially-random mask $\mathcal{M}_j$ to give a spatially-random image $I_j(x,y)$ recorded by a position-sensitive detector $P$. Repeating this process for $N$ different masks gives an ensemble of random intensity maps $\{I_j(x,y)\}$, where $j=1,2,\cdots,N$. (b) Recording of corresponding ensemble of bucket signals $\{B_j\}$, in the presence of a sample $O$, using a position-insensitive detector (``bucket'' detector) $B$. The entrance surface $W$ of the sample coincides with the plane occupied by $P$.}
    \label{fig:Generic_GI_setup}
\end{figure}

A {\it  computational} imaging variant of ghost imaging was later developed \cite{Shapiro2008computational}. See Fig.~\ref{fig:Generic_GI_setup}. Here, spatially-random illumination patterns are produced by a mask that is pre-measured (see Fig.~\ref{fig:Generic_GI_setup}(a)) or otherwise known, and hence do not need to be measured during the ghost imaging procedure.  With the illumination patterns known, bucket signals may then be measured as shown in Fig.~\ref{fig:Generic_GI_setup}(b).  Here, the resolution of the final ghost image is limited to the resolution to which the mask is characterized. If a mask can be characterized to a superior resolution than conventional imaging (e.g., using another probe), super-resolution imaging can be achieved. The {\it computational} ghost imaging variant is extremely similar to the single-pixel camera concept \cite{duarte2008single,sun2016singlePixel}, however, here the illumination is patterned rather than the detector.  This is an important distinction when considering darkfield imaging techniques as well as minimizing dose incident on the object. Further information is given in several review articles \cite{Erkmen2010,shih2012physics,padgett2017introduction}.  We also note the strong similarity between computational ghost imaging using random illumination patterns, and a time-of-flight spectroscopy technique applying pseudo-random chopper or spin-flipper sequences to neutron beams; this latter method may be interpreted as a temporal form of neutron ghost imaging, since it cross-correlates the resulting data with the applied random sequence to yield neutron time-of-flight spectra \cite{NeutronTemporalGI1,NeutronTemporalGI2,NeutronTemporalGI3,NeutronTemporalGI4,NeutronTemporalGI5,NeutronTemporalGI6}. 

Moving beyond the domain of visible-light optics, ghost imaging has now been realized using hard x-rays \cite{Yu2016fourier,pelliccia2016experimental,Schori2017xray}, ultra-cold atom beams \cite{Khakimov2017ghost,Hodgman2019} and electrons \cite{li2018electron}. Furthermore, a proposal exists for Fourier-transform neutron ghost imaging utilizing the anti-bunching property of fermionic fields \cite{NeutronGIproposal}, as a novel means to image micro-magnetic and related structures.   One particular motivation, for further work in both hard x-ray and neutron variants of ghost imaging, is that the penetrating power of such radiation and matter wave-fields will enable tomographic variants of ghost imaging.  In this context, the experimental proof of concept for x-ray ghost tomography has recently been achieved \cite{KingstonOptica2018, KingstonIEEE2019}. The suite of existing neutron-imaging methods \cite{NeutronOpticsHandbook, NeutronImagingAndItsApplications} may, in future, be usefully extended via the addition of neutron ghost imaging.  In this context, note that ghost imaging may be considered to be a generalization of conventional pixel-wise imaging paradigms, at least for the classical and computational variants of the method, since conventional scanning probe (or point-wise) imaging corresponds to a particular choice of needle-like illumination e.g.~via a scanned pinhole \cite{gureyev2018}. Another consideration is that high-resolution neutron microscopy is notoriously difficult (conventional imaging is currently limited to a resolution of about 10$\mu$m \cite{trtik2015improving}), with research being conducted into efficient scintillator materials \cite{trtik2015isotopically} to reduce scintillator thickness (and thus increase spatial resolution) without sacrificing detected neutron flux. Neutron microscopy through compound refractive lenses is also being explored \cite{NeutronMicroscope1, NeutronMicroscope2, NeutronCRL, NeutronCRL2}, and while promising, does not yet provide resolutions matching conventional imaging. 

Within the context established above, three broad motivations for pursuing neutron ghost imaging (NGI) may be raised:
\begin{itemize}
    \item NGI enables position sensitivity to be added to a variety of neutron instruments that are not position resolving.  Examples include instruments for triple-axis neutron spectrometry, small-angle neutron scattering, time-of-flight spectrometry, strain scanning and reflectometry. Such an augmentation would appear to be reasonably straightforward, as the later development of this paper shall demonstrate via experimental proof of concept.
    \item NGI gives a simple and readily implementable route to super-resolution by enabling the resolving power (or pixel size) of a given position-sensitive neutron detector to be significantly increased; each pixel becomes a bucket detector used to form a ghost image up to the resolution to which the speckle generating mask is known. Super-resolution by NGI has the potential to enable microscopy and even ultra-microscopy while still employing thick, cheap scintillator screens with a high stopping power.
    \item NGI provides the ability to yield isotope-resolved images via prompt-%and delayed-%
    gamma-ray bucket detection.  A dark-field version of the method is also possible, in which the bucket detector records neutrons scattered through an appreciable angle.  %The above-mentioned possibilities give both motivation and context for the neutron ghost-imaging work reported below.
    \end{itemize} 

We close this introduction by summarizing the remainder of the paper.  Section~\ref{Sec:Background} reviews some of the general background for ghost imaging.  This section also establishes a protocol for neutron ghost imaging.  Section~\ref{Sec:Methods} describes the experimental methods used to obtain the results in Sec.~\ref{sec:Results}.  Computational neutron ghost imaging and super-resolution via computational neutron ghost imaging, are separately treated in Secs.~\ref{sec:ResultsA} and \ref{sec:ResultsB}.  We discuss some implications of our results in Sec.~\ref{sec:Discussion},  followed by some potential future applications in Sec.~\ref{sec:Future}.  We conclude with Sec.~\ref{sec:Conclusion}.

\section{Background}\label{Sec:Background}

Here we outline key aspects of computational neutron ghost imaging.  We draw on generic background developed in visible-light studies: see e.g.~the previously-cited review articles \cite{Erkmen2010,shih2012physics,padgett2017introduction} and references therein.  We also draw on protocols developed for ghost imaging using hard x-rays \cite{Yu2016fourier,pelliccia2016experimental,Schori2017xray,zhang2018table,pelliccia2018towards,KingstonOptica2018,KingstonIEEE2019}.  This latter link arises from the fact that the connection between neutron ghost imaging and hard x-ray ghost imaging is necessarily close, since (1) both are highly penetrating illumination probes, for samples opaque to visible light, electrons, atomic beams, molecular beams etc.; (2) thermal neutrons have a de Broglie wavelength on the order of $10^{-10}$~m, similar to the wavelength of hard x-rays; (3) while neutron sources are typically significantly less brilliant than corresponding x-ray sources, source sizes can be made comparable, and experimental imaging geometries often have similar spatial dimensions; (4) both x-ray and neutron optics often employ optical elements that are similar in nature---e.g.~crystal beamsplitters, compound refractive lenses, scintillator-coupled position-sensitive detectors etc.---and are  qualitatively different to their visible-light counterparts; (5) efficient high-resolution spatial light modulators, which are readily available for optical studies with visible-light, and form a key component of many computational ghost-imaging setups, have yet to be realized for both neutron and x-ray optics.

\begin{figure}
    \centering
    \includegraphics[width=0.95\linewidth]{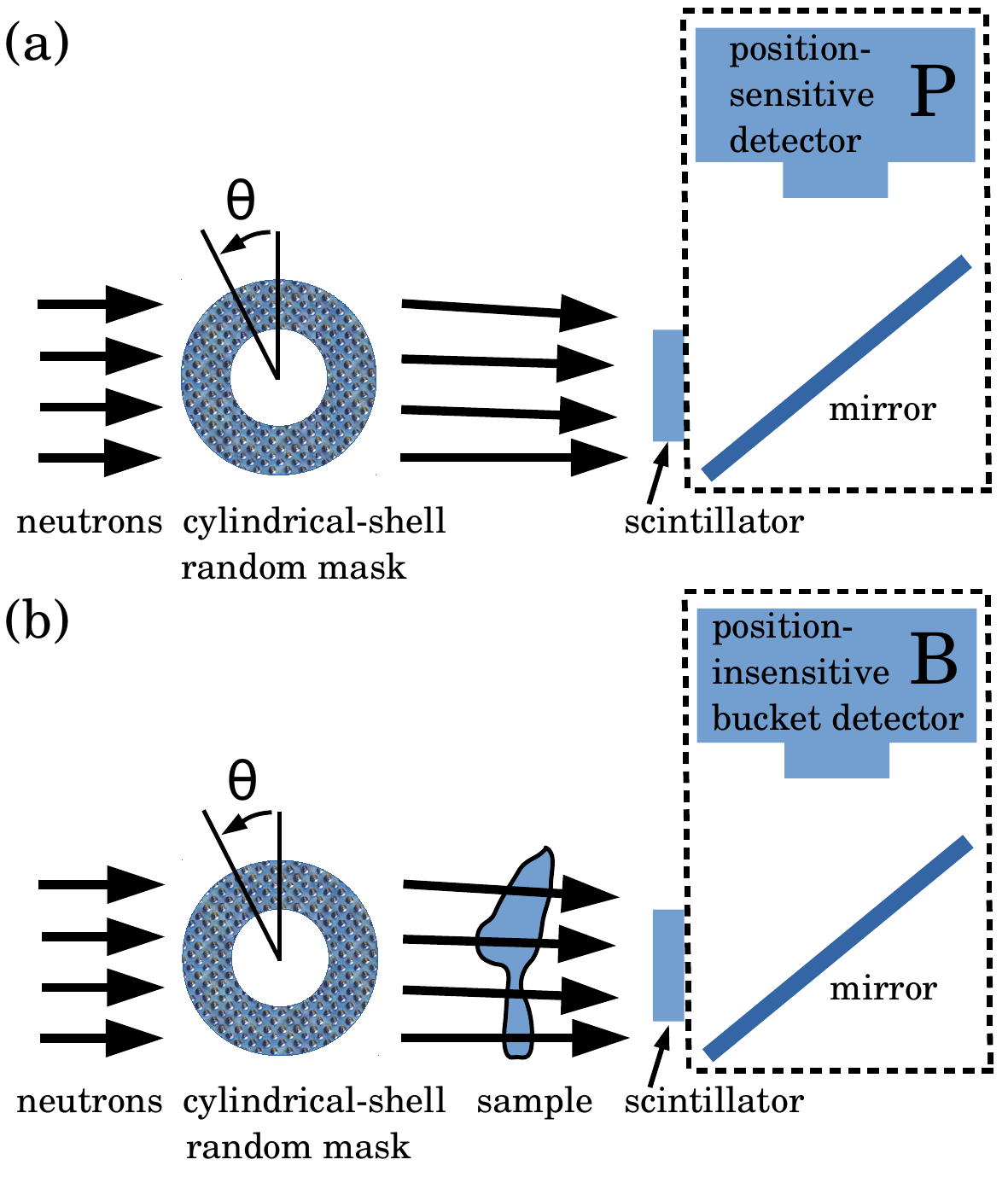}
    \caption{Schematic for computational neutron ghost imaging.  (a) Recording of spatially-random illumination patterns, in the absence of a sample, using a position sensitive detector $P$. (b) Recording of bucket signals, in the presence of a sample, using a position-insensitive detector (``bucket'' detector $B$).}
    \label{fig:GeneralExperimentSchematic}
\end{figure}

Consider an ensemble of $N$ spatially-random intensity distributions $\{I_j(x,y)\}$, where $j=1,\cdots,N$ and $(x,y)$ are transverse Cartesian coordinates in planes orthogonal to an optical axis $z$.  We speak of these distributions as ``speckle'' maps, in a more general usage of the term than that which equates ``speckle'' with ``fully developed coherent speckle''.  The ensemble of random intensity maps may be generated as shown in Fig.~\ref{fig:GeneralExperimentSchematic}(a).  Here, a spatially-uniform beam of $z$-directed neutrons illuminates a speckle-generating mask composed of a cylinder that comprises either (1) a thin cylindrical shell whose surface is coated with a spatially-random distribution of highly-neutron-absorbent particles such as gadolinium oxysulfide (Gadox) powder, or (2) a pair of thin cylindrical shells between which is contained a spatially-random highly-neutron-absorbent material such as randomly-packed steel ball bearings or sodium chloride grains.  Rotating this illuminated mask through a series of azimuthal orientations $\theta$ about its axis $A$, as well as displacing it parallel to $A$, generates the required ensemble $\{I_j(x,y)\}$.  This ensemble may be (1) measured once and for all using the position-sensitive detector $P$, or (2) may be computationally inferred if the three-dimensional (3D) structure of the mask has been accurately characterized (e.g.~using neutron or x-ray tomography) and the properties of the illuminating neutron beam (divergence, spectrum etc.) are well known and stable. The speckles should have high contrast $\kappa$, since the signal-to-noise ratio (SNR) of the resulting ghost image is proportional to $\kappa$ \cite{PaganinOneMask}. Assume that the mask is constructed from a granular material with average grain diameter ({i.e.}, speckle width), $w$. The range of $w$ that can be utilized for a given ghost imaging resolution, $\phi$, is limited. For $w \le \phi$ contrast and resolution arise from the presence/absence of grains; insufficient contrast may result when $w \ll \phi$ or when grain packing is too dense. For $w > \phi$ the resolution is dictated by the sharpness of grain edges, however, $w \gg \phi$ would cause these edges to be sparse; the number of speckle positions required must increase accordingly.
%Since the speckle width, $w$, determines the spatial resolution of the ghost image \cite{ferri2010differential, pelliccia2018towards} (cf.~\citeauthor{NeutronTemporalGI1} \cite{NeutronTemporalGI1}), $w$ should be selected to match the required spatial resolution.

%AMK: I'm not sure that I agree with this David... why can't I have speckle made from large grains but with sharp edges, e.g., large bearings? I would have thought that the sharp edges would enable a sharp ghost image regardless of speckle/grain size?
%DMP: I agree that this needs to be reworded. How about we define w as the characteristic transverse length scale of the spatially random intensity distribution that is produced by the mask?    

The actual ghost-imaging experiment is shown in Fig.~\ref{fig:GeneralExperimentSchematic}(b).  Here, the previously-known ensemble of illuminations $\{I_j(x,y)\}$ impinges upon a thin sample with intensity transmission function $T(x,y)$, where $0 \le T(x,y) \le 1$. Assuming unit efficiency for simplicity, the so-called bucket signal $B_j$ measured by a large single-pixel detector for the $j$th illumination pattern is
\begin{equation}\label{eq:BucketCoefficients}
    B_j=\iint I_j(x,y) \, T(x,y) \, dx \, dy.
\end{equation}
The inverse problem \cite{Sabatier1990,Sabatier2000} of computational ghost imaging then seeks to reconstruct $T(x,y)$ given $\{I_j(x,y),B_j\}$. The cross-correlation (XC) method \cite{katz2009compressive,Bromberg2009ghost} approximates the sample's transmission function via 
\begin{equation}\label{eq:XC_GI_formula}
    T(x,y) \otimes \textrm{PSF}(x,y)\equiv \frac{1}{N}\sum_{j=1}^{N} (B_j-\overline{B})I_j(x,y),
\end{equation}
where $\overline{B}=\textrm{E}(B_j)$, $\textrm{E}$ denotes expectation value, $\otimes$ denotes two-dimensional convolution and $\textrm{PSF}(x,y)$ is the effective point spread function (PSF) associated with the ghost-imaging reconstruction.  This PSF is given by the autocovariance of the ensemble of speckle maps \cite{ferri2010differential,pelliccia2018towards} (cf.~\citeauthor{NeutronTemporalGI1} \cite{NeutronTemporalGI1}):
\begin{equation}\label{eq:PSF_via_autocovariance}
  \textrm{PSF}(x-x',y-y')=\frac{\mathcal{N}}{N}\sum_{j=1}^N I_j(x,y) \, I_j(x',y').    
\end{equation}
Here, $\mathcal{N}$ is a normalization constant chosen such that the PSF integrates to unity, and the assumption of spatial stationarity allows us to the express the left side as a function of coordinate differences $(x-x',y-y')$.  An improved estimate can be obtained by applying Landweber iteration to the XC formula in Eq.~(\ref{eq:XC_GI_formula}), to give an iterative cross-correlation method (IXC) that has a narrower PSF.  See \citeauthor{pelliccia2018towards} \cite{pelliccia2018towards} and \citeauthor{KingstonOptica2018} \cite{KingstonOptica2018,KingstonIEEE2019}, together with references therein, for details regarding IXC ghost imaging.  The narrower PSF arising from IXC has the cost of increased reconstruction noise, this being the usual tradeoff between noise and spatial resolution \cite{GureyevNRU}. IXC reconstructions may be improved via suitable regularization that incorporates constraints such as sparsity in image space, sparsity in image-gradient space etc.~\cite{KingstonOptica2018,KingstonIEEE2019}.

\section{Experimental methods}\label{Sec:Methods}

Experiments were performed using the open-pool reactor-based neutron source on the DINGO imaging beamline at the Australian Centre for Neutron Scattering (ACNS) \cite{Garbe2015,Garbe2017}.  An unfiltered poly-energetic neutron beam was employed, with a spectrum corresponding to thermal neutrons having maximum spectral intensity at wavelength $1.5$\AA.
%$1.08$\AA~\cite{Garbe2015}, and a full-width at tenth maximum of 2.8\AA.
The detector consisted of a 6LiF/ZnS:Cu scintillation screen of thickness 50$\mu$m, a mirror, and an Teledyne Photometrics Iris 15 sCMOS camera placed out of the neutron beam. The sCMOS camera has a $2960 \times 5056$ pixel array with a pixel pitch of 25.7$\mu$m. The detector was positioned $L=9.8$m from a $d=9.8$mm pin-hole at the neutron source giving a beam divergence \cite{Treimer2009} of $\Theta = d/L= 1/1000$. A two-section 5m long flight-tube filled with He at ambient pressure was used to reduce neutron scatter from air. In this configuration, the brightness of the neutron radiation was $9.0 \times 10^6 \textrm{n}. \textrm{cm}^{-2} \textrm{s}^{-1}$.
%Brightness $5.3 \times 10^7 \textrm{n}. \textrm{cm}^{-2} \textrm{s}^{-1}$ when $L/D=500$.

For this set of experiments, the rotation stage used for tomography experiments on the DINGO beamline was used to vary the speckled illumination from the mask. Therefore, a cylindrical mask was employed as depicted in Fig.~\ref{fig:GeneralExperimentSchematic}. The cylindrical mask used for generating the speckle images was placed 150mm upstream from the detector on an Aerotech ABRS 250 air-bearing rotation stage. Potential masks demonstrated to date have been formed from layers of granular materials such as metallic powders \cite{Song2017}, and sand \cite{Kim2013}; we note that foams (the inverse of grains) could also be used, with the extreme 2D case being a stencil. The mask used here consisted of grains of iodized table salt (NaCl), with an average diameter of $1.3$mm. The salt grains were placed inside concentric aluminium cylinders with 1mm thick walls, an inner diameter of 40mm, and an outer diameter of 60mm. Note that the inner diameter is larger than the field-of-view (FOV) of ghost imaging; this ensures that the speckle properties (such as total transmission and magnification) are approximately constant so that the speckle images appear as two sets of granular layers translated in opposite directions.

Here we are performing computational ghost imaging, the first step of which is to record a set of high-quality images of the speckle illumination patterns produced by the salt grains. This step is depicted in Fig.~\ref{fig:GeneralExperimentSchematic}(a). The mask was rotated to 1716 different positions, $\theta$, with an angular increment of $\Delta \theta = 0.21$ degrees and the set of illumination patterns generated was measured with 40 second exposure time. The second step is to record the bucket data, i.e., the total interaction (transmission in this case) of the object with each illumination pattern recorded in step 1. This step is depicted in Fig.~\ref{fig:GeneralExperimentSchematic}(b). The object was placed in the beam in contact with the scintillation screen and the set of illumination patterns are repeated from step 1. In this case, the bucket detector $B$ was generated through software-binning of data recorded with the same detector as in step 1, i.e., position sensitive detector $P$. In step 2, the set of mask positions was rapidly repeated with 5 second exposure time.

We present four experiments.  The first demonstrates computational ghost imaging (CGI) with neutrons.  The remaining three experiments explore the use of CGI to achieve super-resolution images. Two objects were used for the experiments: (1) A cadmium (Cd) stencil constructed from a 400$\mu$m thick sheet of Cd with three holes drilled with diameters of 1mm, 3mm, and 5mm. (2) A resolution star of diameter 20mm with 128 radial lines of width 1.4 degrees created using laser ablation on a gadolinium (Gd) sputtered glass substrate \cite{SiemensStarNeutrons2007}.

{\it Ghost imaging:} The first experiment imaged the Cd stencil object and generated bucket measurements by software-binning a $100 \times 100$ pixel subset of the recorded data to simulate a $2.57 \times 2.57$mm$^2$ bucket. The subset contained the 1mm hole in the stencil.

{\it Super-resolution:} The three super-resolution experiments all image a $13.16 \times 13.16$mm$^2$ FOV assuming position sensitive detectors of decreasing pixel pitch as follows: $0.822$mm, $411\mu$m, and $206\mu$m. The FOV is then captured as $8 \times 8$, $16 \times 16$, and $32 \time 32$ pixel images. We consider each pixel as a bucket detector and refer to these low-resolution images as 2D arrays of bucket detectors. Super-resolution is achieved by performing ghost imaging on a per pixel (or bucket detector) basis. The second experiment again images the Cd stencil while the last two experiments image a quadrant of the resolution star.

\section{Results}\label{sec:Results}

The results will be presented in three subsections. Firstly the speckle illumination patterns generated by the salt grain mask will be analyzed to determine the upper and lower limits to ghost imaging resolution that can be expected. Secondly, computational ghost imaging (CGI) will be demonstrated with neutrons. Lastly, the proposed method to achieve neutron super-resolution imaging using CGI on a per-pixel basis will be demonstrated.

\begin{figure*}
    \centering
    \begin{minipage}{0.24\linewidth}
    \centering
    \scriptsize{(a)}\\[4ex]
    \includegraphics[width=1.0\linewidth]{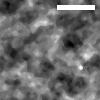}
    \end{minipage}%
    \begin{minipage}{0.36\linewidth}
    \centering
    \scriptsize{(b)}\\
    \includegraphics[width=1.0\linewidth]{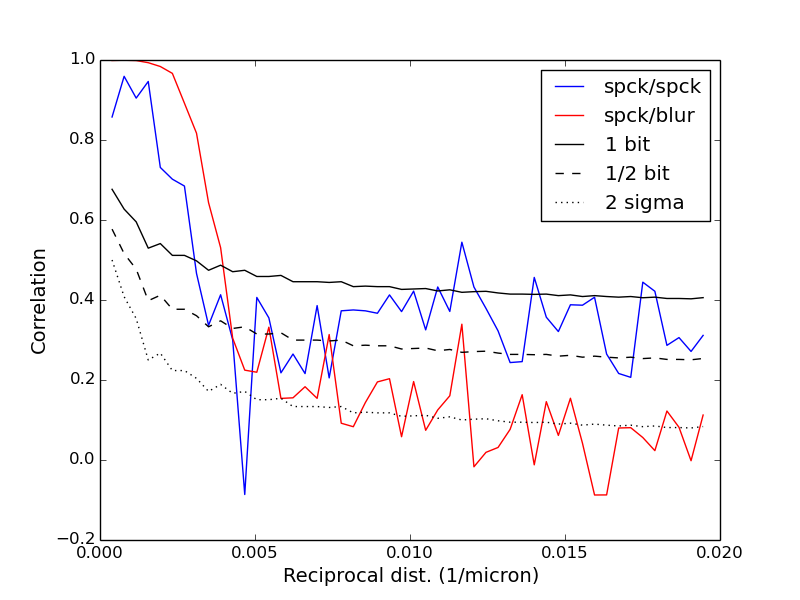}
    \end{minipage}%
    \begin{minipage}{0.36\linewidth}
    \centering
    \scriptsize{(c)}\\
    \includegraphics[width=1.0\linewidth]{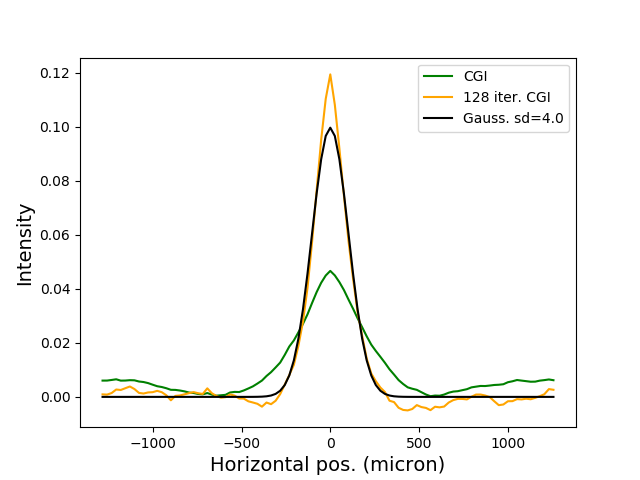}
    \end{minipage}
    \caption{(a) Example $100 \times 100$ pixel region of the speckle intensity image generated by salt. 1mm scale bar. (b) Speckle resolution analysis using Fourier ring correlation. Correlation of repeated measurements (blue plot, labeled {\it spck/spck}) contains a similar amount of information to a measurement correlated to itself after blurring by a Gaussian with a standard deviation of 103$\mu$m (red plot, labeled {\it spck/blur}). (c) Profiles through the PSF generated by CGI (green) and 128 Landweber iterations of CGI (orange) cf.~Gaussian function with a standard deviation of 103$\mu$m (black).}
    \label{fig:speckle}
\end{figure*}

\subsection{Speckle analysis}\label{sec:ResultsSpeckle}

All experiments utilized the same pre-recorded set of high-quality speckle illumination patterns for computational ghost imaging and super-resolution. An example $100 \times 100$ pixel region of this speckled illumination is presented in Fig.~\ref{fig:speckle}(a). The contrast of these speckle images according to a form of Michelson visibility is $\kappa = 0.31$. The calculation adopted here was that of Eq.~(57) in \cite{PaganinOneMask} that modified Michelson visibility to be less sensitive to extreme values such as those from detected gamma rays in a neutron imaging context; the equation is derived in footnote 53 of \cite{PaganinOneMask}.

Based on the Shannon--Nyquist sampling theorem, the upper limit to ghost imaging resolution is twice the pixel pitch of the speckle images; $51.4\mu$m in this case. However, this limit may not be reached if either (1) the PSF of the imaging system used to record the speckle patterns degrades resolution, or (2) the sharpness of the speckle generated by the mask does not attain this spatial frequency. A better estimate for the upper limit to resolution can be achieved by analyzing the resolution of the speckle images themselves. A common technique for estimating image resolution is that of Fourier ring correlation (FRC) \cite{van1987similarity}. Here the correlation of the information in two images is plotted as a function of spatial frequency. This has been presented as the blue curve in Fig.~\ref{fig:speckle}(b) for repeated measurements of the speckle presented in Fig.~\ref{fig:speckle}(a). The resolution is estimated as 0.004$\mu$m$^{-1}$. Adopting the Houston criterion \cite{houston1926fine} for resolution this corresponds to a full-width at half-maximum (FWHM) of $250\mu$m and equates to a Gaussian PSF with a standard deviation of 103$\mu$m. The FRC result of a speckle image compared with itself blurred by this Gaussian function is presented as the red curve in Fig.~\ref{fig:speckle}(b) and displays similar resolution.

The lower limit to ghost imaging resolution is characterized by the PSF generated by the autocovariance of the ensemble of speckle maps as defined in Eq.~(\ref{eq:PSF_via_autocovariance}). A profile through the PSF of the central pixel is presented as the green curve in Fig.~\ref{fig:speckle}(c). Again defining resolution as the FWHM of the PSF, this equates to $463\mu$m for the salt grains. Refining the PSF though the 128 Landweber iterations of the cross-correlation algorithm results in the orange profile shown in Fig.~\ref{fig:speckle}(c) and demonstrates that, in the noise-free case, the upper limit to resolution from FRC analysis (black curve) can be attained.

\begin{figure*}
    \centering
    \begin{minipage}{0.16\linewidth}
    \scriptsize{~}
    \end{minipage}
    \hfill
    \begin{minipage}{0.16\linewidth}
    \centering
    \scriptsize{(i) Conventional image of the sample, i.e., bucket-pixel resolution}
    \end{minipage}
    \hfill
    \begin{minipage}{0.16\linewidth}
    \centering
    \scriptsize{(ii) Standard CGI}
    \end{minipage}
    \hfill
    \begin{minipage}{0.16\linewidth}
    \centering
    \scriptsize{(iii) 128 Landweber CGI iterations}
    \end{minipage}
    \hfill
    \begin{minipage}{0.16\linewidth}
    \centering
    \scriptsize{(iv) 128 regularized Landweber CGI iterations}
    \end{minipage}
    \hfill
    \begin{minipage}{0.16\linewidth}
    \centering
    \scriptsize{(v) High-resolution image of sample, i.e., target image}
    \end{minipage}\\
    \begin{minipage}{0.16\linewidth}
    \flushleft{
        \scriptsize{
            \begin{center}
                Experiment (a)\\[2ex]
            \end{center}
            Sample: Cd stencil\\
            Bucket array: $1 \times 1$\\
            Bucket pitch: 2.57mm.\\
            Image array: $100 \times 100$\\
            Pixel pitch: 25.7$\mu$m.\\
            Zoom factor: 100
        }
    }
    \end{minipage}
    \hfill
    \begin{minipage}{0.16\linewidth}
    \centering
    \scriptsize{(a-i)}\\
    \includegraphics[width=1.0\linewidth]{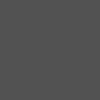}
    \end{minipage}
    \hfill
    \begin{minipage}{0.16\linewidth}
    \centering
    \scriptsize{(a-ii)}\\
    \includegraphics[width=1.0\linewidth]{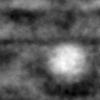}
    \end{minipage}
    \hfill
    \begin{minipage}{0.16\linewidth}
    \centering
    \scriptsize{(a-iii)}\\
    \includegraphics[width=1.0\linewidth]{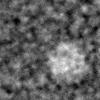}
    \end{minipage}
    \hfill
    \begin{minipage}{0.16\linewidth}
    \centering
    \scriptsize{(a-iv)}\\
    \includegraphics[width=1.0\linewidth]{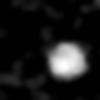}
    \end{minipage}
    \hfill
    \begin{minipage}{0.16\linewidth}
    \centering
    \scriptsize{(a-v)}\\
    \includegraphics[width=1.0\linewidth]{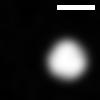}
    \end{minipage}\\
    \begin{minipage}{0.16\linewidth}
    \flushleft{
        \scriptsize{
            \begin{center}
                Experiment (b)\\[2ex]
            \end{center}
            Sample: Cd stencil\\
            Bucket array: $8 \times 8$\\
            Bucket pitch: 1.65mm.\\
            Image array: $256 \times 256$\\
            Pixel pitch: 51.4$\mu$m.\\
            Zoom factor: 32
        }
    }
    \end{minipage}
    \hfill
    \begin{minipage}{0.16\linewidth}
    \centering
    \scriptsize{(b-i)}\\
    \includegraphics[width=1.0\linewidth]{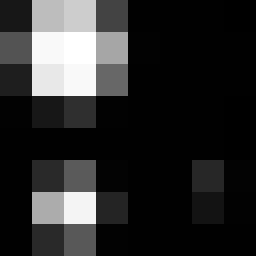}
    \end{minipage}
    \hfill
    \begin{minipage}{0.16\linewidth}
    \centering
    \scriptsize{(b-ii)}\\
    \includegraphics[width=1.0\linewidth]{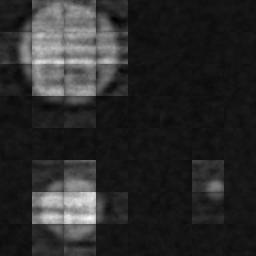}
    \end{minipage}
    \hfill
    \begin{minipage}{0.16\linewidth}
    \centering
    \scriptsize{(b-iii)}\\
    \includegraphics[width=1.0\linewidth]{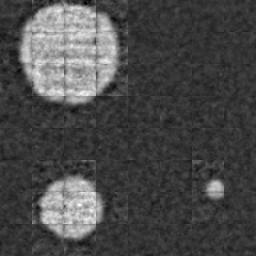}
    \end{minipage}
    \hfill
    \begin{minipage}{0.16\linewidth}
    \centering
    \scriptsize{(b-iv)}\\
    \includegraphics[width=1.0\linewidth]{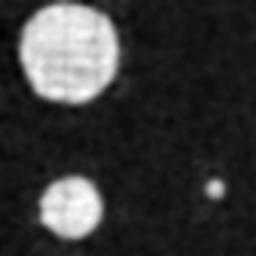}
    \end{minipage}
    \hfill
    \begin{minipage}{0.16\linewidth}
    \centering
    \scriptsize{(b-v)}\\
    \includegraphics[width=1.0\linewidth]{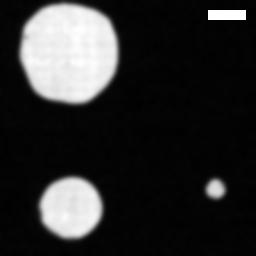}
    \end{minipage}\\
    \begin{minipage}{0.16\linewidth}
    \flushleft{
        \scriptsize{
            \begin{center}
                Experiment (c)\\[2ex]
            \end{center}
            Sample: Res. star\\
            Bucket array: $16 \times 16$\\
            Bucket pitch: 0.82mm.\\
            Image array: $256 \times 256$\\
            Pixel pitch: 51.4$\mu$m.\\
            Zoom factor: 16
        }
    }
    \end{minipage}
    \hfill
    \begin{minipage}{0.16\linewidth}
    \centering
    \scriptsize{(c-i)}\\
    \includegraphics[width=1.0\linewidth]{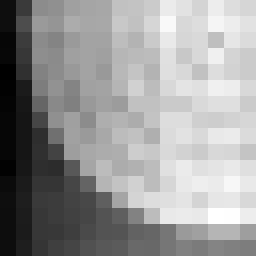}
    \end{minipage}
    \hfill
    \begin{minipage}{0.16\linewidth}
    \centering
    \scriptsize{(c-ii)}\\
    \includegraphics[width=1.0\linewidth]{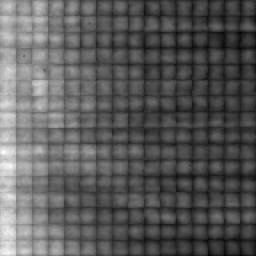}
    \end{minipage}
    \hfill
    \begin{minipage}{0.16\linewidth}
    \centering
    \scriptsize{(c-iii)}\\
    \includegraphics[width=1.0\linewidth]{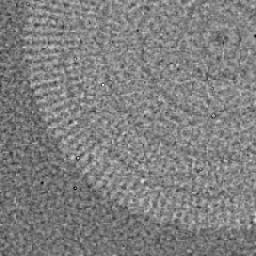}
    \end{minipage}
    \hfill
    \begin{minipage}{0.16\linewidth}
    \centering
    \scriptsize{(c-iv)}\\
    \includegraphics[width=1.0\linewidth]{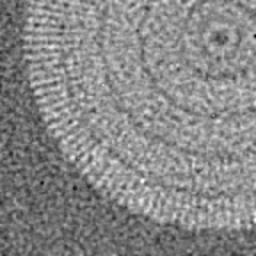}
    \end{minipage}
    \hfill
    \begin{minipage}{0.16\linewidth}
    \centering
    \scriptsize{(c-v)}\\
    \includegraphics[width=1.0\linewidth]{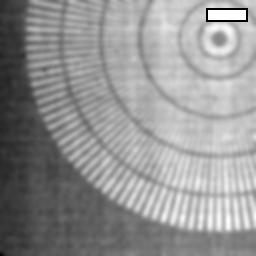}
    \end{minipage}\\
    \begin{minipage}{0.16\linewidth}
    \flushleft{
        \scriptsize{
            \begin{center}
                Experiment (d)\\[2ex]
            \end{center}
            Sample: Res. star\\
            Bucket array: $32 \times 32$\\
            Bucket pitch: 0.41mm.\\
            Image array: $256 \times 256$\\
            Pixel pitch: 51.4$\mu$m.\\
            Zoom factor: 8
        }
    }
    \end{minipage}
    \hfill
    \begin{minipage}{0.16\linewidth}
    \centering
    \scriptsize{(d-i)}\\
    \includegraphics[width=1.0\linewidth]{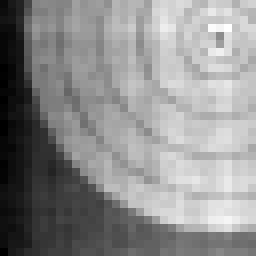}
    \end{minipage}
    \hfill
    \begin{minipage}{0.16\linewidth}
    \centering
    \scriptsize{(d-ii)}\\
    \includegraphics[width=1.0\linewidth]{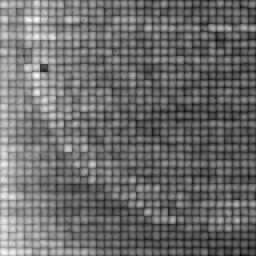}
    \end{minipage}
    \hfill
    \begin{minipage}{0.16\linewidth}
    \centering
    \scriptsize{(d-iii)}\\
    \includegraphics[width=1.0\linewidth]{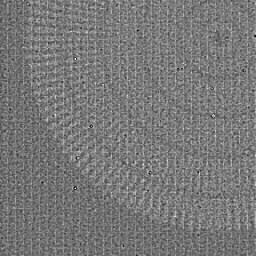}
    \end{minipage}
    \hfill
    \begin{minipage}{0.16\linewidth}
    \centering
    \scriptsize{(d-iv)}\\
    \includegraphics[width=1.0\linewidth]{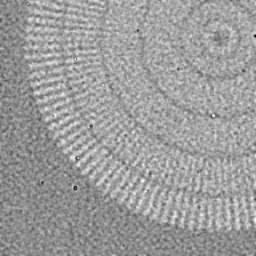}
    \end{minipage}
    \hfill
    \begin{minipage}{0.16\linewidth}
    \centering
    \scriptsize{(d-v)}\\
    \includegraphics[width=1.0\linewidth]{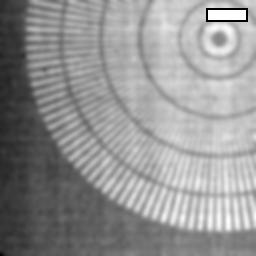}
    \end{minipage}
    \caption{Results for (a) ghost imaging (1mm scale bar) and (b--d) super-resolution by ghost imaging (2mm scale bar). The details for each experiment are given in the left panel. The data recorded for each experiment were collected with resolution as displayed in column (i) referred to as a {\it bucket array} with a pixel or {\it bucket pitch} as specified. In all cases 1716 speckle-image and bucket-measurement pairs were used, with speckle images ``pre-recorded'' or ``known'' at the specified image resolution. The results computed by various CGI methods (ii--iv) are all at the same resolution (that of the speckle images), however, have required different {\it zoom factors} to achieve this (as specified). The right panel show the samples imaged at this same resolution, providing the target image in each case.}
    \label{fig:resultTable}
\end{figure*}

\subsection{Computational ghost imaging}\label{sec:ResultsA}

In order to demonstrate computational ghost imaging (CGI), the 1mm hole drilled into the Cd stencil was imaged. A conventional image of the stencil blurred by a Gaussian PSF with a standard deviation of 103$\mu$m is presented in Fig.~\ref{fig:resultTable}(a-v). Given that 1716 speckle patterns and bucket values were collected, a ghost image of reasonable quality could be expected up to $42 \times 42$ pixels. However, being a stencil image, several strong priors could be asserted on the image generated that compensates for missing measurements. As a result a $100 \times 100$ pixel ghost image was achievable. The results of image recovery by (a) conventional CGI, i.e., cross-correlation, (b) 128 Landweber iterations of CGI, and (c) 128 regularized iterations of CGI (assuming sparsity in image-gradient space) are presented in Fig.~\ref{fig:resultTable}(a-ii)--\ref{fig:resultTable}(a-iv).

\subsection{Super-resolution by computational ghost imaging}\label{sec:ResultsB}

For this demonstration however, $512 \times 512$ pixel speckled illumination patterns were simply recorded by conventional imaging with a $25.7\mu$m pixel pitch, i.e., with a FOV of $13.2$mm. Given the FRC results in Sec.~\ref{sec:ResultsSpeckle}, that speckle illumination resolution is $250\mu$m, these images were binned to $256 \times 256$ pixels with a $51.4\mu$m pixel pitch. Note that the lines at the boundary of the resolution star are spaced at $245\mu$m intervals, right at the resolution limit. We show three super-resolution scenarios given three different coarse detectors with 1.645mm, 0.822mm, and 0.411 pixel pitch. We will demonstrate the potential of CGI for super-resolution, by treating each coarse pixel as a bucket detector and, given the speckle illumination patterns within each coarse pixel, perform CGI per pixel to generate finer sampled images with a pixel pitch matching the speckle pattern images, namely $51.4\mu$m. 

{\it Cadmium stencil:} Here we assume a small $8 \times 8$ pixel camera with pixel pitch of 1.645mm to cover the full 13.2mm FOV. The object is the Cd stencil with 5mm, 3mm, and 1mm holes; a conventional image of the object with this coarse detector is given in Fig.~\ref{fig:resultTable}(b-i). The object imaged with a high-resolution camera ($51.4\mu$m pixel pitch) blurred by a Gaussian function with a standard deviation of $103\mu$m is presented in Fig.~\ref{fig:resultTable}(b-v). The results of super-resolution imaging with a zoom factor of 32 recovered by (a) conventional CGI per bucket pixel, (b) 128 Landweber iterations of CGI per bucket pixel, and (c) 128 regularized iterations (assuming sparsity in image gradient space) are presented in Fig.~\ref{fig:resultTable}(b-ii)--\ref{fig:resultTable}(b-iv).

{\it Resolution star:} The resolution star has been imaged under two scenarios: firstly, using a $16 \times 16$ pixel camera with pixel pitch of 0.822mm to cover the full 13.2mm FOV; secondly, using a $32 \times 32$ pixel camera with a 0.411mm pixel pitch. Conventional images of the object with these coarse detectors are given in Fig.~\ref{fig:resultTable}(c-i) \& \ref{fig:resultTable}(d-i). The object imaged with a high-resolution camera ($51.4\mu$m pixel pitch) blurred by a Gaussian function with a standard deviation of $103\mu$m is presented in Fig.~\ref{fig:resultTable}(c-v) \& \ref{fig:resultTable}(d-v). The results of super-resolution imaging with zoom factors of 16 and 8 respectively recovered by (a) conventional CGI per bucket pixel, (b) 128 Landweber iterations of CGI per bucket pixel are presented in Fig.~ \ref{fig:resultTable}(c-ii)--\ref{fig:resultTable}(c-iii) \& \ref{fig:resultTable}(d-ii)--\ref{fig:resultTable}(d-iii). Observe in these sets of images that artifacts arise on the boundaries of the bucket-pixels. The ghost imaging result is significantly improved after employing a Fourier filtering method (as demonstrated in Fig.~13.39 of \citeauthor{hecht2017optics} \cite{hecht2017optics}) to suppress these bucket-pixel boundary artifacts during 128 regularized iterations of CGI (assuming image smoothness) as presented in Fig.~\ref{fig:resultTable}(c-iv) \& \ref{fig:resultTable}(d-iv).

\section{Discussion}\label{sec:Discussion}

Neutron computational ghost imaging (CGI) has been successfully demonstrated in Fig.~\ref{fig:resultTable}(a) for a 1mm hole in a Cd stencil. Observe that standard CGI by cross-correlation (XC) in Fig.~\ref{fig:resultTable}(a-ii) appears more faithful to the target image (Fig.~\ref{fig:resultTable}(a-v)) than that from 128 Landweber iterations of XC (IXC) in Fig.~\ref{fig:resultTable}(a-iii). This is true since a $100 \times 100$ pixel image has been computed from only 1716 speckle images. The speckle images form a basis for image representation \cite{ceddia2018random,gureyev2018} and 10,000 basis members are required for this image size, if they are orthogonal (more are required in this case). Regularization can compensate for this missing information by using knowledge of the sample properties and making {\it a posteriori} assertions. This has been demonstrated in in Fig.~\ref{fig:resultTable}(a-iv) by assuming sparsity in image-gradient space.

Acquiring 1716 speckle images, and having an average speckle width of $w = 1.3$mm, limits the total FOV that can be achieved by ghost imaging. However, this is perfectly suited to demonstrating super-resolution, replacing one large bucket detector with a small array of {\it bucket pixels} each of which can be treated as a separate ghost imaging experiment to yield a super-resolution image. This concept has been demonstrated in Fig.~\ref{fig:resultTable}(b)--\ref{fig:resultTable}(d) with the images improved from conventional imaging in column (i) improved through CGI to those in column (iv). The Cd stencil experiment in Fig.~\ref{fig:resultTable}(b) demonstrates the dramatic improvement in resolution possible with a zoom factor of 32; this is possible since the stencil is the easiest case for ghost imaging from an SNR perspective (as discussed in Sec.~\ref{Sec:Background}) as well as due to the possibility for more powerful {\it a posteriori} assertions in regularization. For the resolution star experiments: Fig.~\ref{fig:resultTable}(c) shows a significant increase in resolution (zoom factor of 16), but to a lesser extent than the stencil; Fig.~\ref{fig:resultTable}(d) only has a zoom factor of 8 but demonstrates that the expected resolution (as shown in the target image in column (v)) can be achieved.

With each bucket pixel zoomed in to $32 \times 32$ pixels in experiment (b) in Fig.~\ref{fig:resultTable}, $1024$ orthogonal speckle images are required for a complete basis. 1716 non-orthogonal, random, speckle patterns were used and appear to be insufficient, however, regularization can compensate for this lack of information. Only 256 and 64 orthogonal speckle images are required for experiments (c) and (d) respectively in Fig.~\ref{fig:resultTable}. The same 1716 speckle patterns were used in these cases therefore less regularization was required (only low-range image smoothness was assumed). This seemed to be approaching a sufficient set for experiment (d).

Observe that the standard CGI images obtained by XC (column (ii) in Fig.~\ref{fig:resultTable}) for experiments (b--d) contained significant artifacts where the majority of the {\it bucket pixel} contains non-zero intensities; this is again related to the SNR discussion in Sec.~\ref{Sec:Background}. These issues are largely overcome by employing IXC (as demonstrated in column (iii)). The most significant artifacts that remain are those related to bucket pixel boundaries. These are commonly called {\it blocking artifacts} in image processing and are a common issue in super-resolution \cite{van2006image}, image-tiling \cite{hecht2017optics} and image compression \cite{reeve1984reduction} contexts. For the Cd stencil results in Fig.~\ref{fig:resultTable}(b), the heavy regularization possible largely overcame these artifacts, however, for the resolution star images, these had to be explicitly removed by Fourier filtering (as described in Sec.~\ref{sec:ResultsB}) since less regularization was employed.
%what resolution was achieved in 8x8 zoom star? why is this better than speckle analysis? reduced noise levels (refer to T. Gureyev paper \cite{GureyevNRU})
%where could this super-resolution take neutron imaging in the future?

The question of signal-to-noise ratio (SNR) is also worth discussing.  In the high-brilliance limit, the SNR of the XC method in Eq.~(\ref{eq:XC_GI_formula}) is \cite{PaganinOneMask}:
\begin{equation}\label{eq:StencilsAreEasier}
    \textrm{SNR}=\kappa\sqrt{N/n_{\textrm{sample}}}.
\end{equation}
Here $\kappa\in[0,1]$ is the Michelson visibility (i.e.~the contrast) of the ensemble of illuminating speckle fields $\{I_j(x,y)\}$, and $n_{\textrm{sample}}$ is the number of degrees of freedom for the sample transmission function.  For a binary transmission function $T(x,y)$ that only takes the values of zero or unity, $n_{\textrm{sample}}=T_A/a$, where $T_A$ is the area over which $T(x,y)$ is equal to unity, and $a$ is the area occupied by the PSF.  Equation~(\ref{eq:StencilsAreEasier}) quantifies the natural dependencies that (1) the SNR of a neutron ghost image is proportional to the contrast of the masks in the speckle fields from which it is additively composed via Eq.~(\ref{eq:XC_GI_formula}); (2) the SNR is proportional to the square root of the number of utilized masks, a dependence that arises from the random-basis character \cite{Gorban2016,ceddia2018random} of the ensemble of illuminating speckle maps; (3) For fixed $\kappa$ and $N$, the SNR becomes lower as the number of degrees of freedom in $T(x,y)$ becomes larger, consistent with the observation that ``stencil like'' transmission functions (namely those for which $T_A\ll a$) have relatively higher SNR in XC ghost imaging reconstruction when compared to transmission functions for which the inequality $T_A\ll a$ does not hold.  Also, if source brightness $\mathcal{B}$ is taken into account, Eq.~(\ref{eq:StencilsAreEasier}) becomes
\begin{equation}\label{eq:StencilsAreEasier2}
    \textrm{SNR}=\left( \frac{n_{\textrm{sample}}}{\kappa^2N}+\frac{\Xi}{\mathcal{B}} \right)^{-1/2},
\end{equation}
where $\Xi$ is a constant that is proportional to the total exposure time and inversely proportional to the resolution-element area.  In the limit where 
\begin{equation}
    \mathcal{B} \gg \Xi \kappa^2 N / n_{\textrm{sample}},
\end{equation}
Eq.~(\ref{eq:StencilsAreEasier2}) levels out at the high-brilliance asymptote \cite{Erkmen2010} of Eq.~(\ref{eq:StencilsAreEasier}).  In the low-brilliance limit we have $\textrm{SNR} \sim \sqrt{\mathcal{B}}$.  

\section{Potential future applications of neutron ghost imaging}\label{sec:Future}
%Looking beyond the proof-of-concept reported here, what future applications do we envisage for neutron ghost imaging? 
One of the strengths of the ghost imaging (GI) concept is its ability to add spatial resolution to non-spatially resolving measurements. In other words, GI may be a viable alternative to conventional imaging in measurement situations where a pixel-array detector is not available, or not practical. 

\begin{figure}
    \centering
    \includegraphics[width=1.0\linewidth,trim={0 1cm 0 0},clip]{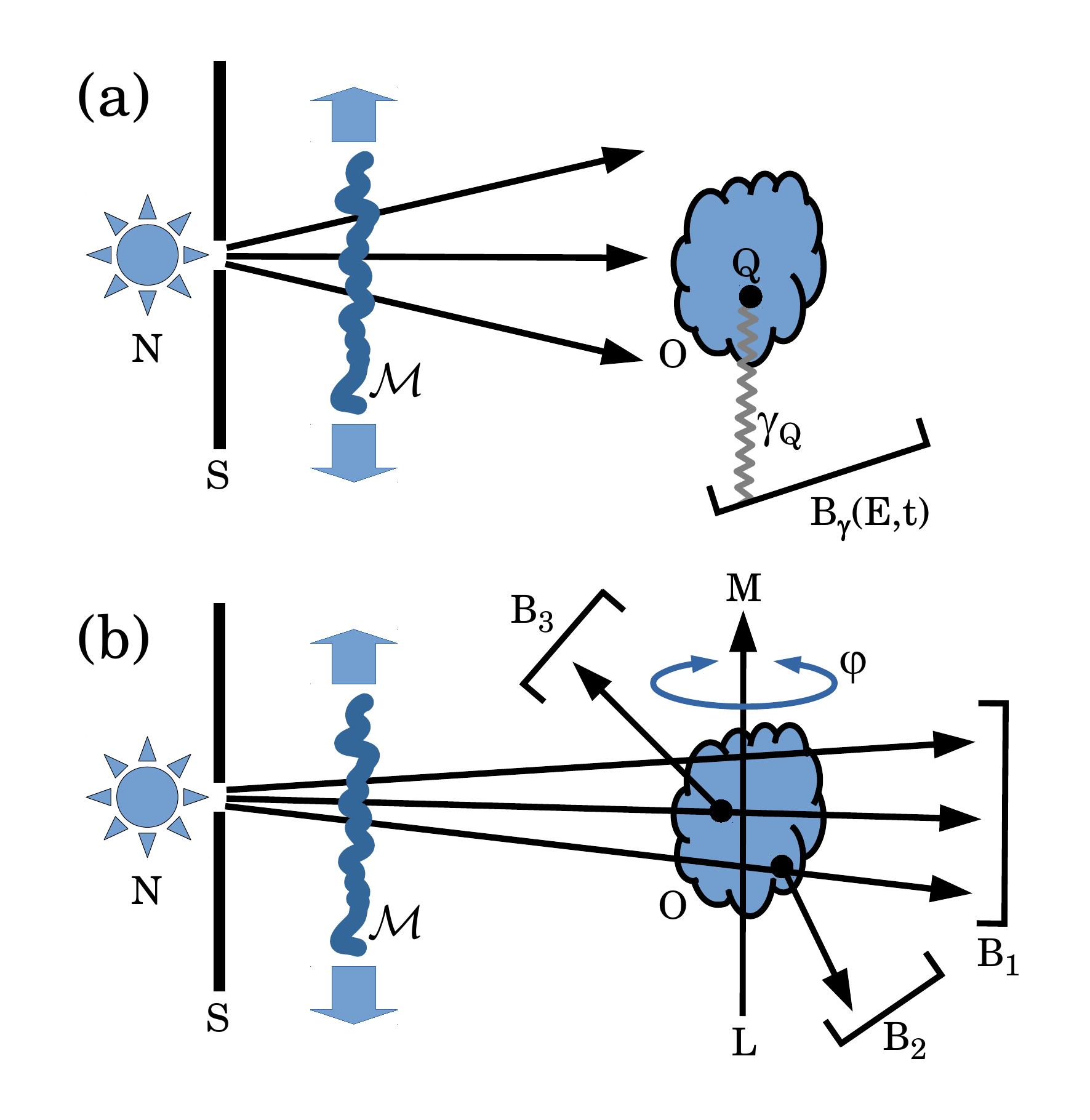}
    \caption{(a) Setup for isotope-resolved neutron ghost imaging using prompt gamma-ray %and delayed-gamma-ray
    detection. For each position of a well-characterized random mask $\mathcal{M}$, points such as $Q$ within the object $O$ may emit neutron-induced prompt-%and delayed-%
    gamma radiation $\gamma_Q$ to give a bucket signal $B_{\gamma}(E,t)$ that is measured using a gamma-ray spectrometer.  This bucket signal may be measured as a function of both gamma-ray energy $E$ and emission time $t$.  (b) Setup for neutron ghost tomography, in which the object $O$ may be rotated about the axis $LM$ to a variety of azimuthal angles $\varphi$. Bucket $B_1$ may be used for bright-field neutron ghost tomography, while buckets $B_2$ and $B_3$ yield dark-field neutron ghost tomograms.}
    \label{fig:future-work}
\end{figure}

One such technique is prompt-gamma neutron activation analysis (PGAA), which measures the elemental composition of a sample through measuring the intensity and energy of prompt gamma rays emitted by a sample that is irradiated by a neutron beam. Gamma spectrometers employed for these measurements could be multiplexed by a GI approach whereby the bucket signal $B_{\gamma}(E,t)$ measured by a gamma-ray spectrometer (see Fig.~\ref{fig:future-work}(a) for a depiction of this process) could be cross-correlated with the illumination to produce a spatially-resolved elemental composition of the sample.

The problem of adding spatial resolution is shifted from the detector to the illumination beam, enabling one to use existing detector technology in a novel fashion. Note that, in the spirit of the present work, the spatial resolution of such measurements could be tailored by adjusting the mask structure (and number of illuminations) so as to suit any specific experiment.

While the present paper has been devoted to neutron ghost imaging in two spatial dimensions, the penetrating power of neutrons enables the technique to be extended to 3D neutron ghost imaging (``ghost tomography''), as recently accomplished with hard x-rays \cite{KingstonOptica2018,KingstonIEEE2019}.  
% For example, neutron ghost tomography could be pursued by adapting protocols recently developed for hard-x-ray ghost tomography \cite{KingstonOptica2018,KingstonIEEE2019}.  
In fact, complementing existing methods for bright-field \cite{Treimer2009} and dark-field \cite{Strobl2008} neutron tomography, one could devise the measurement scheme sketched in Fig.~\ref{fig:future-work}(b), whereby different bucket detectors acquire transmitted and scattered neutrons, while the mask displacement and the sample rotation permits multiple tomograms to be constructed using methods described in \citeauthor{KingstonOptica2018} \cite{KingstonOptica2018,KingstonIEEE2019}.

The two examples given here are meant to illustrate some general guidelines towards augmenting existing neutron techniques with GI. We seek to inspire discussion around these topics, without going into the details of specific techniques. It is important to remark however, that mask design is a topic of intense research (see \citeauthor{higham2018deep} \cite{higham2018deep} for video-rate optical imaging, but also work in hard x-rays by \citeauthor{KingstonIEEE2019} \cite{KingstonIEEE2019}), which will greatly improve the ability to produce illumination masks that are optimized for the sample at hand. The imaging problem is thus recast in terms of beam shaping---avoiding building complexity into the detector design---and computational algorithms, in the spirit of what is broadly described as computational imaging: a hybrid hardware--software imaging system \cite{paganin2004}, able to overcome the limitations of optics and pixel-array detector systems.

\section{Conclusion}\label{sec:Conclusion}

A protocol for computational neutron ghost imaging was outlined, and applied to two separate experiments.  The first achieved computational neutron ghost imaging by illuminating a sample with an ensemble of spatially random neutron fields and subsequently registering the total sample transmission using a single bucket detector.  This enables position resolution to be incorporated into a variety of neutron-scattering instruments, that do not currently possess imaging capability. The second experiment used neutron ghost-imaging concepts to achieve super-resolution.  Here, a ghost image was independently reconstructed for each pixel of a detector with coarse spatial resolution, thereby increasing the effective spatial resolution of the detector.  Avenues for future work were outlined, including tomographic neutron ghost imaging, dark-field neutron ghost imaging, and isotope-resolved color neutron ghost imaging via prompt-
%and delayed-%
gamma-ray bucket detection.

\begin{acknowledgments}
AMK and GRM acknowledge the financial support of the Australian Research Council and FEI-Thermo Fisher Scientific through Linkage Project LP150101040, and the use of supercomputer time provided by Australia’s National Computational Infrastructure (NCI). DP, DMP, and GRM acknowledge travel support from the Australian Nuclear Science and Technology Organisation. The authors acknowledge useful discussions with Jeremy Brown, Margaret Elcombe, Wilfred Fullagar, Tim Petersen, Kirrily Rule, Anton Stampfl and Imants Svalbe.
\end{acknowledgments}

%\appendix
%
%\section{First appendix}
%
%\section{Second appendix}

\bibliography{NeutronGI}% Produces the bibliography via BibTeX.

\end{document}